\documentclass[lettersize,journal]{IEEEtran}
\usepackage{amsmath,amsfonts}
\usepackage{algorithmic}
\usepackage{algorithm}
\usepackage{array}
\usepackage{caption}
\usepackage{xcolor}
\usepackage{textcomp}
\usepackage{stfloats}
\usepackage{url}
\usepackage{verbatim}
\usepackage{graphicx}
\usepackage{cite}
\usepackage{fancyhdr} 
\usepackage{subfigure}
\usepackage{booktabs} 
\usepackage{amssymb}
\usepackage{amsbsy}
\begin{document}

\newcommand{\NAME}[0]{TrafficLLM}

\title{Self-Refined Generative Foundation Models for Wireless Traffic Prediction}

\author{Chengming Hu, Hao Zhou,~\IEEEmembership{Member, IEEE}, Di Wu, Xi Chen, Jun Yan, and Xue Liu, ~\IEEEmembership{Fellow, IEEE}.\vspace{-20pt}
\thanks{Chengming Hu, Hao Zhou, Xi Chen, and Xue Liu are with the School of Computer Science, McGill University, Montreal, QC H3A 0E9, Canada. (e-mail: chengming.hu@mail.mcgill.ca, hao.zhou4@mail.mcgill.ca, xi.chen11@mcgill.ca, xueliu@cs.mcgill.ca).}
\thanks{Di Wu is with the Department of Mechanical, Industrial and Aerospace Engineering, Concordia University, 1515 Ste-Catherine St. W., Montreal, QC H3G 2W1, and Department of Electrical and Computer Engineering, McGill University, Montreal, QC H3A 0E9, Canada. (e-mail: di.wu5@mcgill.ca).}
\thanks{Jun Yan is with the Concordia Institute for Information Systems Engineering, Concordia University, Montreal, QC H3G 1M8, Canada (e-mail: jun.yan@concordia.ca)}
\thanks{Chengming Hu, and Di Wu are the corresponding authors.}}

\maketitle

\thispagestyle{fancy}            
\chead{This paper has been accepted by IEEE Transactions on Vehicular Technology. } 

\renewcommand{\headrulewidth}{1pt}      
\pagestyle{plain}

\begin{abstract}

With a broad range of emerging applications in 6G networks, wireless traffic prediction has become a critical component of network management. 
However, the dynamically shifting distribution of wireless traffic in non-stationary 6G networks presents significant challenges to achieving accurate and stable predictions. 
Motivated by recent advancements in Generative AI (GenAI)-enabled 6G networks, this paper proposes a novel self-refined Large Language Model (LLM) for wireless traffic prediction, namely \NAME{}, through in-context learning without parameter fine-tuning or model training. 
The proposed \NAME{} harnesses the powerful few-shot learning abilities of LLMs to enhance the scalability of traffic prediction in dynamically changing wireless environments. 
Specifically, our proposed \NAME{} embraces an LLM to iteratively refine its predictions through a three-step process: traffic prediction, feedback generation, and prediction refinement. 
Initially, the proposed TrafficLLM conducts traffic predictions using task-specific demonstration prompts.   
Recognizing that LLMs may generate incorrect predictions on the first attempt, this paper designs feedback demonstration prompts to provide multifaceted and valuable feedback related to these initial predictions. 
The validation scheme is further incorporated to systematically enhance the accuracy of mathematical calculations during the feedback generation process.
Following this comprehensive feedback, our proposed \NAME{} introduces refinement demonstration prompts, enabling the same LLM to further refine its predictions and thereby enhance prediction performance.
Evaluations on two realistic datasets demonstrate that the proposed \NAME{} outperforms LLM-based in-context learning methods, achieving performance improvements of 23.17\% and 17.09\%, respectively.

\end{abstract}

\section{Introduction}



The envisioned 6G networks are expected to support a wide range of emerging applications, including Vehicle-to-Vehicle (V2V) and Vehicle-to-Infrastructure (V2I) communications, integrated satellite-aerial-terrestrial networks, and integrated sensing and communication, among others. 
The evolving network architectures and highly integrated network functions significantly increase the complexity of network management. 
Specifically, wireless traffic prediction is a crucial aspect of network management, encompassing tasks such as load balancing, energy saving, and more.

Traditional traffic prediction studies commonly apply statistical and time-series analysis approaches, such as Autoregressive Integrated Moving Average (ARIMA)~\cite{kumar2015short}. 
Deep Learning (DL) methods have also demonstrated success in traffic prediction, including Long Short-Term Memory units (LSTM)~\cite{9128037}, and Convolutional Convolution Network~\cite{10265809}, among others, which exhibit considerable advancements over traditional statistical approaches.
However, in non-stationary wireless networks, traffic distributions can dynamically shift over time and across various base stations, Unmanned Aerial Vehicles (UAVs), High Altitude Platforms (HAPs), and diverse end-user devices, resulting in complex and non-linear spatial-temporal characteristics. 
These dynamic shifts in traffic distribution present significant challenges in developing robust conventional DL models capable of adapting to such variations.  
A straightforward strategy is to directly deploy well-trained DL models to unseen traffic; however, this results in overfitting to the pre-training dataset and degraded performance on unseen data. Another common method is to fine-tune DL models on new data, leading to additional computational resources.


Generative AI (GenAI) has attracted considerable attention due to its ability to analyze complex data distributions and generate analogous content. 
The promising features have spurred the exploration of GenAI-enabled wireless networks through the deployment of powerful generative foundation models, particularly Large Language Models (LLMs).
Fu et al.~\cite{10533857} utilized an LLM to design a hybrid network Intrusion Detection System in the Internet of Vehicles (IoV), encompassing a semantic extractor, input embedding, LLM pre-training and fine-tuning.
Liu et al.~\cite{10520918} enhanced energy-efficient and reliable Reconfigurable Intelligent Surface-based IoV communication systems by fine-tuning an LLM on critical IoV data. 
However, the fine-tuning of LLMs is computationally expensive and memory-intensive, posing significant challenges for deploying these LLMs in resource-limited wireless networks~\cite{10685369}. 
Furthermore, large-scale traffic datasets are often scarce in wireless networks. The high measurement cost of packet aggregation compromises packet forwarding performance, making it difficult to accurately measure flow-level traffic at each timestamp. 

While LLMs were originally designed for natural language processing and generation, LLMs have recently demonstrated remarkable effectiveness in numerical time series prediction and mathematical calculation. 
For instance, Wu et al.~\cite{wu2024mathchat} introduced a conversational problem-solving framework for mathematical calculation problems, in which a user proxy agent initiates conversations with an LLM agent by sending mathematical problems along with predefined demonstration prompts. During the interaction, the user proxy agent executes the code generated by the LLM agent and returns the results. The LLM agent continues the problem-solving process until a predefined pattern signals the end of the conversation. 
Imani et al. employed the zero-shot chain-of-thought prompting technique (namely MathPrompter~\cite{imani2023mathprompter}) to enhance the confidence level of mathematical calculation results. Given a mathematical problem, MathPrompter first generates a corresponding algebraic expression, and then designs algebraic and Python prompts that encourage the LLM to solve the problem in different ways. The compute verification step is eventually introduced to validate the solutions with higher confidence. 
Zong et al.~\cite{zong2023solving} designed a series of mathematical problem descriptions along with correct and incorrect example responses as input to an LLM, guiding the model to generate correct mathematical answers by following the correct examples and avoiding incorrect ones. 
Gruver et al.~\cite{gruver2023large} represented time series data as a string of numerical digits and framed time series prediction as a next-token prediction task similar to text generation. By extending discrete distributions to continuous densities capable of modelling complex multimodal patterns, LLMs can effectively assign likelihoods to entire sentences of time series data across various tasks in a zero-shot manner.
Rather than fine-tuning LLMs, in-context learning leverages their inference capabilities by integrating task-specific demonstration prompts to guide LLMs in task execution~\cite{brown2020language}.  
The demonstration prompts convert task-specific examples into well-structured natural language sentences and incorporate multi-step reasoning enhancements, such as chain-of-thought prompting~\cite{wei2022chain}. 
By following contextual examples within demonstration prompts, LLMs can identify the type of task using existing knowledge from pre-training data and subsequently acquire task-solving strategies~\cite{10356715}, enabling them to effectively recognize and learn new tasks.

Given the high computational efficiency without model fine-tuning, in-context learning presents a promising method for addressing various tasks in wireless networks.
Specifically, prompt tuning~\cite{pmlr-v202-phang23a} enables general-domain LLMs to execute wireless network tasks by providing task-specific requirements and objectives through tailored prompts.    
Jiang et al.~\cite{10670195} designed an LLM-based knowledge base that performs semantic extraction by providing the LLM with a personalized prompt base (including character profiles), enabling the model to fulfill specific task requirements. However, such LLM-based knowledge base lacks an iterative self-refinement scheme to validate the reliability of the LLM’s outputs, which can result in incorrect semantic extraction. 
By augmenting LLMs with modular capabilities, LLM-based agent systems~\cite{xi2025rise} are designed to support long-term planning and decision-making, with LLMs serving as intelligent entities capable of learning, reasoning, and executing actions in wireless networks. 
For example, Tong et al. introduced an LLM-based agent (namely WirelessAgent~\cite{tong2025wirelessagent}), which integrates four modules to manage network slicing tasks, including perception, memory, planning, and action. Although an improvement scheme is incorporated in the planning module, WirelessAgent focuses solely on overall bandwidth utilization without considering output format, slicing method, and other performance factors. This prevents WirelessAgent from providing comprehensive feedback on the LLM’s outputs. 
Jiang et al. introduced an LLM-enabled multi-agent system ~\cite{10638533} for semantic communication, which comprises three modules: multi-agent data retrieval, multi-agent collaborative planning, and multi-agent evaluation and reflection. However, multiple LLM-based agents are employed to evaluate and refine the solutions during their interactions, which can lead to additional computational and communicational costs, particularly in bandwidth-limited environments.

To this end, this paper proposes a novel \textbf{\NAME{}}, a self-refined LLM designed for wireless traffic prediction, which leverages in-context learning without parameter fine-tuning or model training. 
Specifically, our proposed \NAME{} employs an LLM that iteratively enhances its predictions through a three-step process: traffic prediction, feedback generation, and prediction refinement. 
The proposed \NAME{} initially incorporates task-specific prediction demonstration prompts, which guide an LLM (e.g., GPT-4~\cite{achiam2023gpt}) in traffic prediction by adhering to contextual instructions and examples embedded within these demonstration prompts. 
Recognizing that LLMs may generate incorrect predictions, this paper designs feedback demonstration prompts to facilitate a thorough evaluation of these predictions. 
These prompts encompass multiple aspects, including prediction performance, prediction format, prediction method, and actionable steps, providing multifaceted and valuable feedback associated with predictions. 
The validation scheme is further incorporated to systematically enhance the accuracy of mathematical calculations during the feedback generation process. 
Following this comprehensive feedback, our proposed \NAME{} introduces refinement demonstration prompts that enable the same LLM to iteratively refine predictions and enhance prediction performance.

In summary, the \textbf{main contributions} of this paper include:
\begin{itemize}
\item We propose a novel \NAME{} that leverages in-context learning for wireless traffic prediction without parameter fine-tuning or model training. 
\item We design task-specific feedback and refinement prompts that guide the LLM in automatically improving its predictions through a three-step process, including traffic prediction, feedback generation, and prediction refinement.
\item Comprehensive experiments on two realistic datasets demonstrate the effectiveness and scalability of our proposed \NAME{} over in-context learning methods. 
\end{itemize}
\section{Self-Refined LLM for Traffic Prediction}

\begin{figure*}[t]
    \centering
    \includegraphics[width=.98\textwidth]{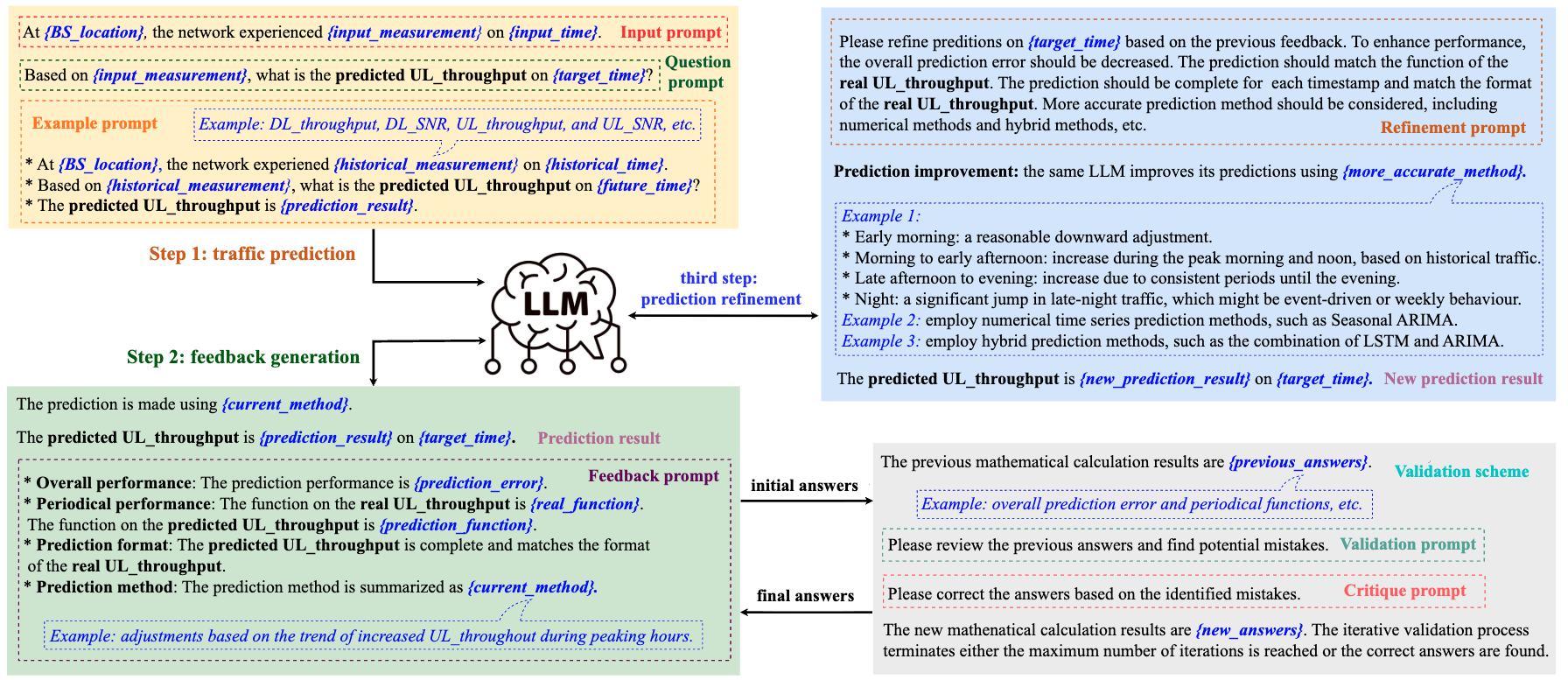}
    \caption{An illustrated example of the proposed \NAME{} applied to uplink throughput (i.e., UL\_throughput) prediction.
    The example prompt presents how to formulate UL\_throughput prediction in a question-answer format. Given the input prompt that includes base station location and network measurements for input time, an LLM is expected to predict future UL\_throughput for the target time by responding to question outlined in the question prompt.
    The proposed \NAME{} automatically generates the feedback prompt by querying the LLM with multiple questions related to the prediction performance. 
    The validation scheme is designed to identify potential inaccuracies during feedback generation and correct incorrect answers. 
    The refinement prompt is further incorporated to allow the same LLM to revise its own predictions by providing actionable instructions.}\label{fig:method}
    \vspace{-3mm}
\end{figure*}

\subsection{Problem Formulation}

Given a wireless network, a sample of the network measurements $r[t]=[r_1[t],\cdots,r_{N}[t];t]$ is recorded every $24/T$ hour, where $r_1[t]$ indicates the network traffic and $r_n[t] (n\ge 2)$ represents other network factors at timestep $t$, such as base station location, and downlink throughput, among others. 
For time-series wireless traffic prediction, a sliding window of $w+l$ samples are applied, where the first $w$ samples construct the input $x[t]$, and the traffic load of the remaining $l$ samples constitute the ground truth $y[t]$. Concretely, $x[t]$ and $y[t]$ are represented as follows:
\begin{equation}
\begin{aligned}
x[t]&=[r_1[t],\dots,r_1[t+w-1],\dots,r_N[t], \\ 
&\dots r_N[t+w-1];t,\dots,t+w-1],
\end{aligned}
\end{equation}
\begin{equation}
y[t]=[r_1[t+w],\dots,r_1[t+w+l-1]].
\end{equation}
For ease of presentation, we denote $d[t] = \{x[t]; y[t]\}$ as a transformed data sample. The dataset is then generated by consecutively shifting the window forward to create subsequent samples $d[t+1], d[t+2],\dots$, among others. To this end, our traffic prediction problem is formulated as follows:

Given the historical network measurement samples $\{x[t],x[t+1],\dots\}$, our objective is to develop a traffic prediction model that accurately forecasts future traffic samples, closely matching the ground truth $\{y[t],y[t+1],\dots\}$.

\subsection{Traffic Prediction}

The proposed \NAME{} utilizes a pre-trained LLM for wireless traffic prediction through in-context learning, without parameter fine-tuning or model training. 
One critical step involves transforming numerical traffic into natural language sentences, enabling LLMs to leverage inherent pattern analysis capabilities to generate predictions~\cite{10685369}.  
To this end, this paper achieves the data-to-text transformation by following the template-based description~\cite{xue2022translating}. Specifically, the proposed \NAME{} formulates the traffic prediction task in a question-answer manner, where the input prompt $p_{\text{input}}$ provides historical network measurements and network topology and the question prompt $p_{\text{ques}}$ involves inquiries about future traffic.

Fig.~\ref{fig:method} illustrates an example of predicting future uplink throughput (i.e., UL\_throughput) at the target time using our proposed TrafficLLM. 
Leveraging the success of LLMs in recognizing and performing new tasks through contextual information, the task-specific example prompt $p_{\text{exam}}$ is first introduced to guide an LLM in traffic prediction based on historical network measurements and network topology, including base station location (i.e., BS\_location), historical GPS time (i.e., historical\_time), signal-to-interference-and-noise-ratio for both downlink and uplink channels (i.e., DL\_SNR and UL\_SNR), among others.  Subsequently, by adhering to the question-answer format specified in $p_{\text{exam}}$, an LLM $\mathcal{M}(\cdot)$ predicts future traffic $\hat{y}[t]$ for the target time as follows:
\begin{equation}
\hat{y}[t]=\mathcal{M}(p_{\text{exam}}\oplus p_{\text{input}}\oplus p_{\text{ques}}), \label{eqn:traffic_prediction}
\end{equation}
where $\oplus$ denotes the concatenation operation.

\subsection{Feedback Generation}

Recognizing that an LLM may not be able to generate precise predictions on its first attempt, our proposed \NAME{} further incorporates feedback demonstration prompts designed to provide multifaceted and valuable feedback related to these initial predictions. 
The proposed \NAME{} automatically generates feedback demonstration prompts by prompting the LLM with multiple questions, thereby reducing human efforts and enhancing cost efficiency in performance evaluation.
As shown in Fig.~\ref{fig:method}, our feedback demonstration prompt is designed to encompass prediction performance, prediction format, prediction method, and actionable steps:

\begin{itemize}
\item \textbf{Overall performance}: Given ground truths and predictions, the proposed \NAME{} is guided to automatically calculate the overall performance by prompting the same LLM with a question such as, \textit{``What is the Mean Absolute Error of the predictions?}  The overall performance is iteratively improved until convergence is achieved.

\item \textbf{Periodical performance}: Given that time-series data can be projected onto sine and cosine functions in the real domain~\cite{esling2012time}, the proposed \NAME{} automatically represents ground truths and predictions using these functions by prompting a question such as, \textit{``For ground truths and predictions, what are their projected functions derived from the combination of sine and cosine functions?''}
By aligning the projected function of predictions as closely as possible with that of ground truths, the predictions are expected to accurately capture the periodicity of ground truths, including variations during peak traffic hours.

\item \textbf{Prediction format}: The predictions should match the format of the ground truths and be complete for each timestamp. Thus, we prompt the LLM to evaluate the prediction format using a question such as, \textit{``Do the predictions align with the format of the ground truths and provide a complete prediction for each timestamp?''}

\item \textbf{Prediction method}:
To further enhance performance, one actionable step is to encourage the proposed \NAME{} to adopt more accurate methods in future iterations. Thus, the prediction method is summarized as one aspect of the feedback demonstration prompt at each iteration by posing the LLM with a question such as \textit{``What is the prediction method applied in the current iteration?''}
\end{itemize}

Given that the LLM may provide incorrect answers during the feedback generation process, such as errors in calculating overall prediction performance and projected functions, this paper further introduces a validation scheme to systematically enhance the accuracy of mathematical calculations within the generated feedback, as shown in Fig.~\ref{fig:method}. 
Specifically, the same LLM is instructed to self-verify its mathematical answers using the validation prompt, such as \textit{``Please review the previous answers and find potential mistakes''} and subsequently refines the answers based on the critique prompt, for example, \textit{``Please correct the answers based on the identified mistakes''}. 
Thus, our validation scheme identifies mathematical calculation errors during the feedback generation process and produces improved answers conditioned on the critique.
Notably, the validation process is iteratively conducted until either the maximum number of iterations is reached or the final answers are obtained when the answers converge within a small threshold (e.g., when the MAE difference between two consecutive iterations is less than 0.001). 
Our future work will focus on developing a more dedicated validation scheme leveraging chain-of-verification~\cite{dhuliawala2024chain} to enhance reliability.


\subsection{Prediction refinement}

To this end, given thorough feedback demonstration prompts, our proposed \NAME{} additionally incorporates the refinement demonstration prompt to enable the same LLM to refine its predictions. 
As shown in Fig.~\ref{fig:method}, we prompt the same LLM with instructions such as: \textit{``Please refine predictions based on the previous thorough feedback. To enhance performance, more accurate time series prediction methods should be considered, including numerical methods, machine learning methods, and hybrid methods. The prediction should match the function 
of the real UL\_throughput. The prediction should be complete and match the format of the real UL\_throughput.''} 
To this end, the refined prediction $\hat{y}_{i+1}[t]$ is represented as follows:
\begin{equation}
\hat{y}_{i+1}[t]=\mathcal{M}(x[t]\oplus \hat{y}_{i}[t]\oplus p_{\text{feed},i}\oplus p_{{\text{refine}},i}),
\end{equation}
where $p_{\text{feed},i}$ and $p_{{\text{refine}},i}$ are the feedback and refinement demonstration prompts specifically associated with the prediction $\hat{y}_{i}[t]$ at the $i$-th iteration. 
By adhering to these task-specific instructions, the LLM can employ numerical methods (e.g., Seasonal ARIMA) or hybrid methods (e.g., the combination of LSTM and ARIMA) to further refine its predictions.
Furthermore, we retain the history of previous predictions, feedback, and refinements, enabling the LLM to learn from past mistakes and avoid repeating them. Thus, $\hat{y}_{i+1}[t]$ can be further instantiated as follows:
\begin{equation}
\begin{aligned}
    \hat{y}_{i+1}[t]&=\mathcal{M}(x[t]\oplus \hat{y}_{0}[t]\oplus p_{\text{feed},0}\oplus p_{{\text{refine}},0}\oplus \dots \\
    &\oplus \hat{y}_{i}[t] \oplus p_{\text{feed},i}\oplus p_{{\text{refine}},i}). \label{eqn:refine}
\end{aligned}
\end{equation}
Alg.~\ref{alg} summarizes our proposed \NAME{}, including traffic prediction, iterative feedback generation and prediction refinement. Note that \NAME{} is eventually evaluated without feedback and refinement demonstration prompts.

\begin{algorithm}[t]
\caption{\NAME{} for Wireless Traffic Prediction}\label{alg}
\begin{algorithmic}[1]
\STATE \textbf{Require:} Input $x[t]$ and LLM $\mathcal{M}(\cdot)$
\STATE Predict traffic according to Eqn.~(\ref{eqn:traffic_prediction});
\WHILE{not converge}
\FOR{$i=0,1,\dots$}
\STATE Generate feedback $p_{\text{feed},i}$ associated with $\hat{y}_{i}[t]$;
\STATE Validate mathematical calculations within $p_{\text{feed},i}$; 
\STATE Refine predicted traffic $\hat{y}_{i+1}[t]$ according to Eqn.~(\ref{eqn:refine});
\ENDFOR
\ENDWHILE
\STATE \textbf{Return} $\hat{y}_{i+1}[t]$. 
\end{algorithmic}
\end{algorithm}
\section{Experimental Study}

\subsection{Dataset Description}

Our proposed \NAME{} is evaluated on the two following realistic datasets to demonstrate its effectiveness and scalability in wireless networks. 

\subsubsection{V2I Radio Channel Measurement Dataset~\cite{10268872}}
This evaluation dataset captures critical V2I channel measurements simulated through a realistic testbed in the city of Munich, comprising a base station and a terminal installed on a vehicle.  

\subsubsection{Wireless Traffic Dataset~\cite{barlacchi2015multi}}
This evaluation dataset records the traffic load dataset of urban life in the city of Milan. Every 10 minutes, a sample of Internet access interaction is generated, recording the time of user interactions and the base stations managing these interactions.

\subsection{Experiment Setup \& Evaluated Baselines}
This paper performs a one-day-ahead sequence traffic prediction using one-day historical traffic. 
To demonstrate the superiority of our proposed \NAME{} over state-of-the-art methods, we employ GPT-4~\cite{achiam2023gpt} as the foundation model for our proposed \NAME{} and compare its performance against the following baseline methods:
\begin{itemize}

\item \textbf{ARIMA~\cite{kumar2015short}} predicts traffic through a well-known statistical time-series prediction method. 

\item \textbf{LSTM~\cite{9128037}} predicts traffic through multiple LSTM blocks, as a representative of conventional DL methods.

\item \textbf{ST-Tran~\cite{9491035}} predicts traffic through a spatial-temporal transformer, as a representative of more advanced DL methods. Notably, both ST-Tran and LSTM serve as our optimal baselines due to their specific model training and fine-tuning, incurring additional computational cost and memory requirements. 

\item \textbf{GPT-4~\cite{achiam2023gpt}} predicts traffic through solely GPT-4 without feedback-refinement iterations. 

\item \textbf{GPT-3.5~\cite{gpt-3.5}} predicts traffic through solely GPT-3.5 without feedback-refinement iterations. 

\end{itemize}

Note that due to the limited space, the prediction results of fine-tuned LLMs are not included, given their computationally expensive and memory-intensive nature.
Moreover, this paper utilizes \textbf{Mean Absolute Error (MAE)} and \textbf{Mean Square Error (MSE)} as our performance metrics.

\begin{figure}[t]
\centering
\subfigure[Average MAE]{
\includegraphics[width=.22\textwidth]{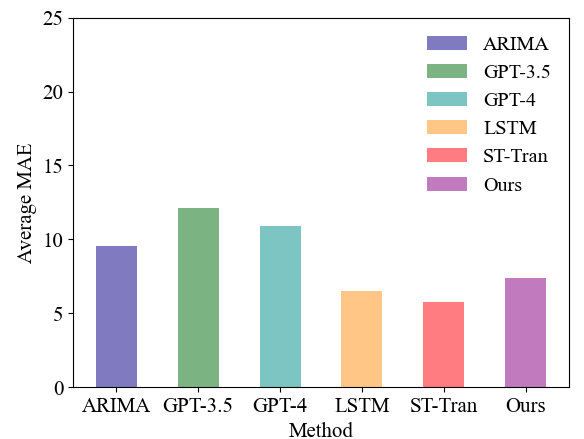} 
}
\subfigure[Average MSE]{
\includegraphics[width=.22\textwidth]{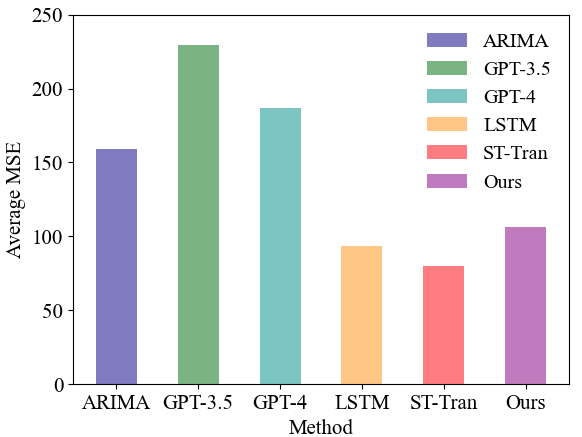}
}
\vspace{-2mm}
\caption{Prediction performance comparison between our proposed \NAME{} and the baselines on the V2I dataset.}
\vspace{-5mm}
\label{fig:pred_performance_same}
\end{figure}

\subsection{Experimental Results}

\subsubsection{Evaluation on the V2I dataset}

We first compare the prediction performance of our proposed \NAME{} against the baselines on the V2I dataset~\cite{10268872}. 
Note that 70\% of the dataset is incorporated with demonstration prompts while the remaining 30\% is utilized for traffic prediction evaluation.  
For the baselines ARIMA and LSTM, the same 70\% of the dataset is chosen to train the models and the remaining data for testing.

Fig.~\ref{fig:pred_performance_same} compares the average MAE and MSE between our proposed \NAME{} and the baselines. 
Although LSTM and ST-Tran surpass our proposed \NAME{} in prediction accuracy, these models incur additional computational costs and substantial memory (e.g., $395\text{s}$ and $11.60\text{MB}$ required for training a five-layer LSTM). 
The additional computational demands make them less practical for traffic prediction on resource-constrained devices in non-stationary wireless networks.
Specifically, when deployed on network devices, our proposed \NAME{} operates solely in the inference phase by accessing GPT-4 via an API. This shifts the computational and memory demands offloaded to the cloud-based infrastructure, allowing efficient operation on resource-constrained devices. 
Although API-based inference may involve certain local memory usage for handling input queries and responses, it eliminates the need to store and update model parameters locally, reducing the overall memory footprint. 
However, deploying conventional DL models on network devices requires storing model parameters locally, as well as local model training and fine-tuning when new, unseen data becomes available on local network devices, leading to memory-intensive demands, particularly when processing large-scale traffic datasets. 
Thus, our proposed \NAME{} is more practical for real-world traffic prediction at network edge devices in non-stationary wireless networks. 
By leveraging cloud-based inference, our proposed \NAME{} provides a more efficient and scalable method in resource-constrained network edge environments.
Given that our proposed TrafficLLM showcases strong generalization across various types of LLMs, one future direction will be to integrate our work with LLM pre-training and fine-tuning, facilitating its potential deployment on powerful cloud servers.

In comparison to LLM-based methods (i.e., GPT-3.5 and GPT-4) and ARIMA, our proposed \NAME{} achieve reductions of up to 39.27\% and 53.62\% in prediction MAE and MSE, respectively. 
Note that our proposed \NAME{} is ultimately evaluated on the testing data without the iterative feedback generation and prediction refinement process. 
We further evaluate the end-to-end latency during the inference process, observing an average latency of 0.2 seconds per testing sample. 
Compared to GPT-4, the proposed \NAME{} improves prediction performance by 23.17\% while maintaining acceptable latency during the inference process, making it practical for real-time applications in wireless networks.
Among GPT-3.5 and GPT-4, one reason for their inferior performance is the absence of our proposed iterative feedback and refinement processes. 
GPT-3.5 assumes that future traffic patterns can closely mirror historical traffic, resulting in predictions that merely replicate historical traffic.
Furthermore, GPT-4 predicts future traffic to more accurately reflect daily traffic patterns, resulting in enhanced performance compared to GPT-3.5. 
Our proposed \NAME{} iteratively improves traffic predictions by incorporating feedback and refinement demonstration prompts. 
This process enables an LLM to utilize accurate prediction methods, such as seasonal ARIMA and ensemble statistical methods, leading to superior performance compared to LLM-based methods and ARIMA.

\subsubsection{Evaluation on the wireless traffic dataset}

\begin{table}[t]
    \centering
    \caption{Performance comparison between our proposed \NAME{} and the baselines on the wireless traffic dataset. }\label{tab:pred_performance_different}
    \begin{tabular}{c|cc}
    \toprule
    Method & Average MAE & Average MSE \\
    \midrule
    LSTM~\cite{9128037} & 10.86 & 154.59 \\
    ST-Tran~\cite{9491035} & 8.31 & 135.27 \\
    GPT-4~\cite{achiam2023gpt} & \underline{14.92} & \underline{216.41} \\
    GPT-3.5~\cite{gpt-3.5} & 17.60  &  299.73   \\
    ARIMA~\cite{kumar2015short} & 19.74 & 302.25 \\
    \midrule
    \textbf{\NAME{}} & \textbf{12.37} ($\color{red}{\Uparrow}$17.09\%) & \textbf{181.22}($\color{red}{\Uparrow}$16.26\%) \\
    \bottomrule
    \end{tabular}
    \vspace{-5mm}
\end{table}

To showcase the scalability of our proposed \NAME{} across dynamically changing wireless networks, this paper compares our \NAME{} against the baselines on the wireless traffic dataset~\cite{barlacchi2015multi}, using traffic from different base stations for demonstration prompts and traffic predictions, respectively.
To this end, we randomly select one base station as the source to integrate its traffic data with demonstration prompts and subsequently choose another base station with a distinct data distribution as the target for traffic prediction evaluation. Note that both baselines LSTM and ST-Tran are pre-trained on the source dataset and then fine-tuned on the target dataset.

Table~\ref{tab:pred_performance_different} demonstrates the average MAE and MSE of our proposed \NAME{} and the baselines. Among the LLM-based prediction methods, the best performance is \textbf{bold}, while the second best is \underline{underlined}. ``$\color{red}{\Uparrow}$'' indicates the performance improvement of our proposed \NAME{} over GPT-4. 
It is observed that all the LLM-based methods, including GPT-3.5, GPT-4, and our proposed \NAME{}, significantly outperform ARIMA, showcasing the strong zero-shot learning capability of LLMs. 
For instance, compared to ARIMA, our proposed \NAME{} achieves reductions in average MAE and MSE by 37.33\% and 40.04\%, respectively. 
Our proposed \NAME{} exhibits notable improvements over GPT-3.5 and GPT-4, with MAE reductions of 29.72\% and 17.09\%, which underscores more powerful generalization capability on unseen datasets through our iterative feedback generation and prediction refinement processes.
Although conventional DL methods (i.e., LSTM and ST-Tran) showcase superior performance, these models require specific pre-training and fine-tuning on the source and target datasets, resulting in additional computational costs and posing significant challenges for resource-constrained devices within wireless networks.
Moreover, due to the dynamically shifting traffic distribution in non-stationary wireless networks, well-trained DL models may overfit the historical traffic used during pre-training, leading to degraded performance on new, unseen traffic over time.

\section{Conclusion}

Motivated by recent advancements of GenAI-enabled wireless networks, this paper proposes a novel \NAME{} designed for wireless traffic prediction through in-context learning without model training or parameter fine-tuning. 
Following our task-specific feedback and refinement demonstration prompts, the proposed \NAME{} iteratively improves traffic prediction within dynamically changing wireless networks. 
Evaluations on two realistic datasets demonstrate the effectiveness and scalability of our proposed \NAME{} over in-context learning methods, with enhanced performance of 23.17\% and 17.09\%, respectively.

\bibliographystyle{IEEEtran}
\bibliography{Reference}
\end{document}